\documentstyle[graphicx]{mn}

\title{High-resolution X-ray imaging and spectroscopy of the core of NGC 4945 with {\it XMM-Newton} and {\it Chandra}}

\author[N.\,J. Schurch, T.\,P Roberts and R.\,S. Warwick]
{N.\,J. Schurch, T.\,P. Roberts \& R.\,S. Warwick\\
Department of Physics and Astronomy, University 
of Leicester, University Road, Leicester, LE1 7RH}
\date{}

\def\asca{{\it ASCA~\/}}

\def\xte{{\it RXTE~\/}}
\def\xmm{{\it XMM-Newton~\/}}

\def\cha{{\it Chandra~\/}}

\def\sax{{\it BeppoSAX~\/}}

\def\H0{{\rm ~km~s^{-1}~Mpc^{-1}}}

\def\etal{et al.~\/}

\def\la{\mathrel{\hbox{\rlap{\hbox{\lower4pt\hbox{$\sim$}}}{\raise2pt\hbox{$<$}}}}}
\def\ga{\mathrel{\hbox{\rlap{\hbox{\lower4pt\hbox{$\sim$}}}{\raise2pt\hbox{$>$}}}}}

\def\d25{D$_{25}$}

\def\.25{0.25 keV\thinspace}

\def\deg{\hbox{$^\circ$}}
\def\arcm{\hbox{$^\prime$}}
\def\arcs{\hbox{$^{\prime\prime}$}}

\begin{document}

\maketitle

\begin{abstract}

We utilize the complimentary capabilities of \xmm and \cha, to conduct
a detailed imaging  and spectral study of the nearby galaxy NGC 4945 
focussing on its nucleus and immediate surroundings (within $\sim 1$ kpc
of the nucleus).
A complex morphology is revealed including a predominantly hard, but 
partially resolved, nuclear source plus a spectrally soft, conically
shaped X-ray ``plume'', which extends $30''$ (500 pc) to the
northwest. In NGC 4945 our direct view of the active galactic 
nucleus (AGN) is blocked below $\sim 10$ keV by extremely heavy 
line-of-sight absorption and the observed X-ray spectrum is dominated
by multi-temperature thermal emission associated with the nuclear 
starburst and the X-ray plume.  Nevertheless the signature of the AGN 
is present in the form of a neutral Compton reflection component
and a 6.4 keV fluorescent iron K$\alpha$ line. We conjecture that the 
site of the continuum reprocessing is the far wall of a highly inclined
molecular torus, a geometry which is consistent with the presence 
of $\rm  H_{2}O$ megamaser emission in this source.
The soft spectrum ($\sim 0.6$ keV) and limb-brightened appearance
of the X-ray  plume  suggest an  interpretation  
in terms of a mass-loaded  superwind emanating from the nuclear starburst.

\vspace{2mm}


\vspace{10mm}

\end{abstract}

\section{Introduction}

NGC  4945 is  a nearby  edge-on  (i $\sim$78$^{o}$;  Ott, 1995)  spiral
galaxy (type SBcd or SABcd), believed  to be a member of the Centaurus
group (Hesser \etal 1984), at a distance  of between 3.7 Mpc and 8.1 Mpc
(here  we  use  3.7  Mpc;  Mauersberger \etal  1996).  Near  infra-red
observations have revealed the nuclear region of NGC 4945 to contain a
powerful, yet  visually obscured  starburst region with  a $\sim$200pc
($\sim$10\arcs)  ring morphology  (Moorwood  \etal 1996)  and a  total
infrared luminosity (8-1000 $\mu$m) of 2.2$\times$10$^{10}$L$_{\odot}$
(Spoon  \etal 2000).  

Although much  of  the central  activity can  be
explained in terms  of a nuclear starburst, the  presence of an active
galactic nucleus (AGN) in NGC 4945 has been confirmed by the detection
of a luminous and variable hard  X-ray source, coincident  with the centre of  
the galaxy (Iwasawa  \etal 1993; Guainazzi  \etal 2000).   NGC 4945  harbours the
brightest known  Seyfert 2 nucleus at  100 keV (Done  \etal 1996), but
extremely  heavy  obscuration cuts-off  the  direct  AGN continuum  at
energies  below $\sim$10  keV  (Guainazzi \etal  2000; Madejski  \etal
2000).  The   measured  X-ray  column   density  ($\sim$4$\times$10$^{24}$ atom
cm$^{-2}$) places it on the threshold of being a Compton-thick Seyfert
2.  After correctly accounting for the effects of Compton scattering in the absorbing 
material (Matt \etal 1999), the inferred intrinsic X-ray (0.1-200 keV) luminosity is 
$\sim$1.8$\times10^{43}$erg s$^{-1}$ (Guainazzi \etal 2000). The  soft  
to medium  energy  X-ray  spectrum  of the  nucleus  is
comprised of both thermal emission associated with a nuclear starburst
and emission  generated through the reprocessing of  the AGN continuum
emission (Guainazzi  \etal 2000).  

The dense clouds  that envelope the
nuclear region produce a rich variety of molecular lines (Curran \etal
2001) including  $\rm  H_{2}O$ megamaser  emission,  which requires  an
edge-on inner disk geometry  (Greenhill, Moran \& Hernstein 1997). The
megamaser  emission provides  a tight  constraint on  the mass  of the
central  black  hole,  $M_{BH}  \approx 1.4  \times  10^{6}  M_\odot$,
implying that  the AGN is radiating  at up to 60\%  of its Eddington
luminosity.

Infrared  and optical
observations of NGC 4945 have  also revealed the presence of a conical
cavity, attributed  to a starburst-driven  superwind, protruding above
the  disk  of   the  galaxy  (Chen  \&  Huang   1997;  Moorwood  \etal
1996).  Despite  heavy obscuration  at  optical wavelengths,  extended
line-emitting  gas,  co-spatial  with  the conical  cavity,  has  been
detected in outflow along the galaxy's minor-axis and is attributed to
the  interaction  between the  superwind  and  the surrounding  medium
(Heckman,  Armus  \&  Miley  1990).  The  low  spatial  resolution  of
the current published data has mitigated against the detection of the superwind in
X-rays, although Guainazzi \etal (2000)  suggest that at least part of
the soft X-ray emission from NGC 4945 may originate in this component.

Here we present new \xmm and  \cha X-ray observations of NGC 4945. The
X-ray imaging  reveals a complex morphology within  the nuclear region
of NGC 4945 and considerable emission associated with the
disk  of   the  galaxy,  including  at  least   12  discrete  X-ray
sources. In  this paper  we will confine  our attention to  a detailed
analysis  of the  nuclear region  of NGC  4945; the  remaining diffuse
emission associated  with the galactic disk and  the off-nuclear point
source population  will be  discussed in  a separate  paper (Roberts,
Schurch \& Warwick 2002; in preparation).
 
\section{Observations and Data Reduction}

\subsection{The \xmm Observation}

NGC 4945  was observed with \xmm  during orbit 205  (2001, January 21)
for $\sim$24 ks as part  of the mission Guaranteed Time programme. The
EPIC MOS and PN instruments were operated in full frame mode, with the
medium  filter, for  the full  duration of  the  observation. Detailed
summaries  of the  \xmm mission  and instrumentation  can be  found in
Jansen  \etal  (2001;  and  references  therein).  The  raw  data  were
processed with the public release version of the \xmm Science Analysis
System  (SAS v 5.2)  standard processing  chains. After  filtering for
data flagged  as bad, X-ray  events corresponding to pattern  0-12 for
the two MOS cameras (similar to  grades 0-4 for \asca) and pattern 0-4
for  the  PN  camera  (single  and  double  pixel  events  only)  were
accepted.  Investigation  of  the  full-field count-rate  revealed  no
significant  background  flaring  episodes  in  the  observation.  The
effective exposure times  for the MOS and PN cameras  were 22.2 ks and
19.2 ks respectively.  Images, spectra  and  corresponding background
information were extracted with the SAS task XMMSELECT.

\subsection{The \cha observation}

The archival  \cha data were obtained  from the UK mirror  of the \cha
X-ray Center  data archive, maintained  by the Leicester  Data Archive
Service (LEDAS).   The \cha observation  itself was performed  on 2000
January 27/28, resulting in a total exposure of 50 ks, with the ACIS-S
array at the prime focus.   Our reduction and analysis of this dataset
was performed using version 2.1 of the CIAO software, and incorporated
many of  the standard analysis procedures available  online from {\tt
cxc.harvard.edu/ciao}.  The data were initially filtered to remove all
events  outside the  0.3-10  keV  energy range.  The  analysis of  the
full-field   light-curve  revealed  the   observation  to   be  heavily
contaminated  by background  flaring.  The  removal of  all  times for
which  the total  detector count  rate exceeded  8 counts  s$^{-1}$ was
found to dramatically reduce the  background in the S3 chip, leaving a
GTI-corrected event-list containing 33.1 ks of ``good'' data.

\section{Joint imaging with {\it CHANDRA} and {\it XMM-NEWTON}}

\xmm  and \cha  images  were  extracted for  a  2\arcm~square  region
centered  on the nucleus  of NGC  4945, revealing  a wealth  of detail
including  a  predominantly  hard  nuclear  source and  a  soft  X-ray
``plume'' extending to the northwest  of the nuclear source. Fig. 1:
{\it Panels  (a)-(c)},  show  high  spatial resolution  energy-coded  X-ray
images from  the \cha ACIS detector,  while Fig. 1: {\it Panel (d)}  
shows an `X-ray
colour' image from the co-added  \xmm EPIC cameras. Whereas the nuclear 
source appears point-like  with {\it XMM-Newton},  the \cha  imaging  partially
resolves  the emission  centred on  the active
nucleus  (Fig, 1: {\it Panel  (c)} ).  The  nuclear  point source  is
revealed  as an extremely  hard source  by both  the \cha  hard band
image and the  \xmm X-ray colour image. This spectral hardness is
almost certainly due to heavy line-of-sight obscuration. Both \xmm and
\cha clearly  resolve the  X-ray plume, establishing  its orientation and 
extent along the minor axis of the galaxy (30\arcs = 500 pc, NW) and 
also its 
intrinsic spectral softness (Fig. 1:  {\it Panels (a),(b),(d)} ). The medium
band \cha image reveals the X-ray plume to have
a limb-brightened morphology, qualitatively similar to the morphology of
the X-ray plume seen in NGC 253 (Strickland \etal 2000).

\begin{figure}
\centering
\begin{minipage}{85 mm} 
\centering
\vbox{

\hbox{
\includegraphics[width=4.1 cm, angle=270]{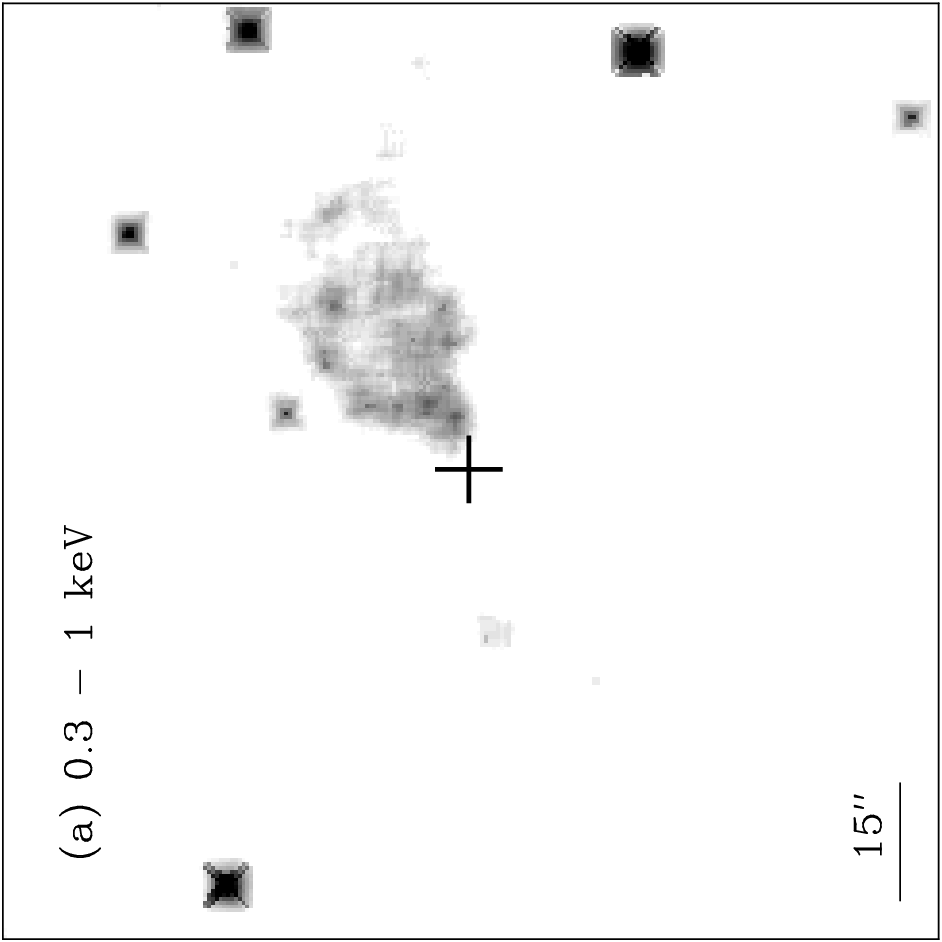}
\includegraphics[width=4.1 cm, angle=270]{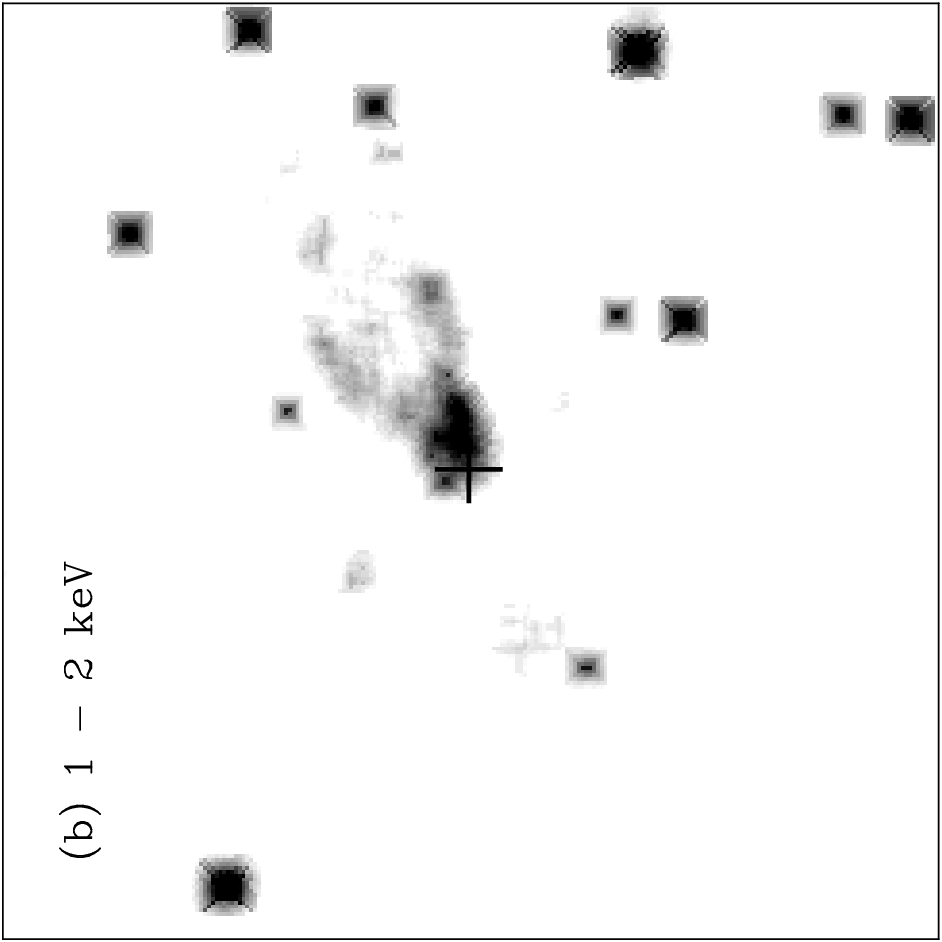}
}

\hbox{
\includegraphics[width=4.1 cm, angle=270]{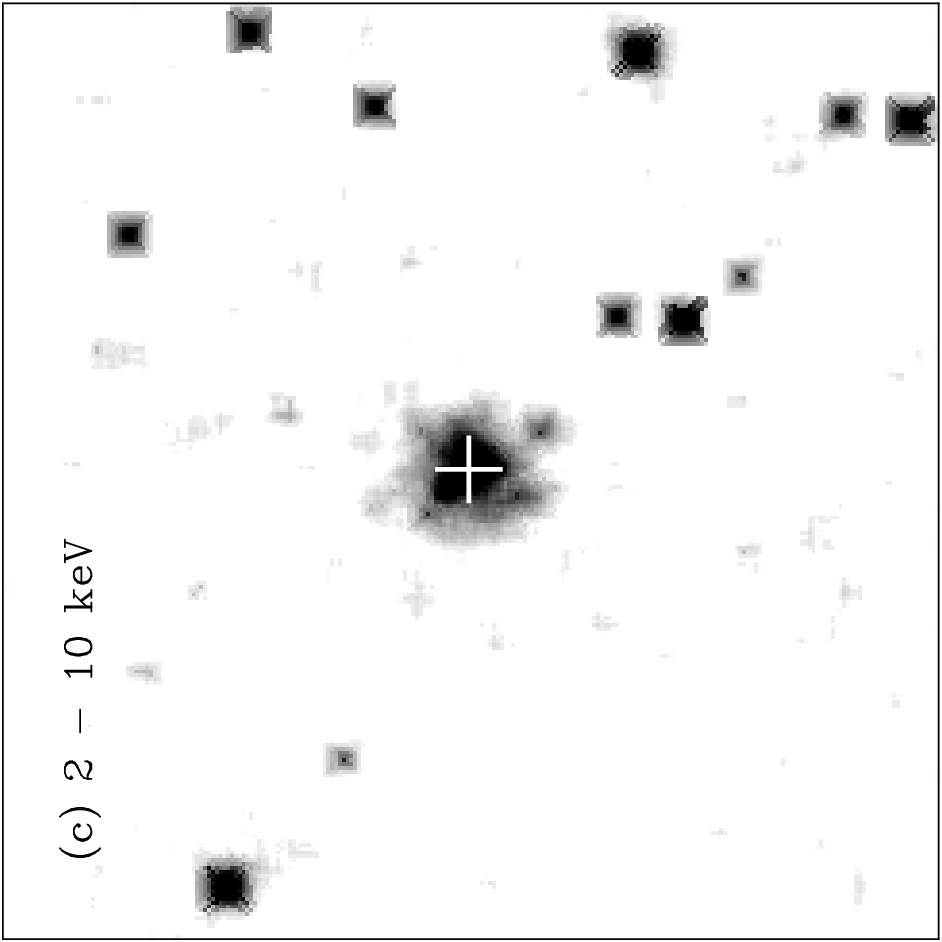}
\includegraphics[width=4.1 cm, angle=270]{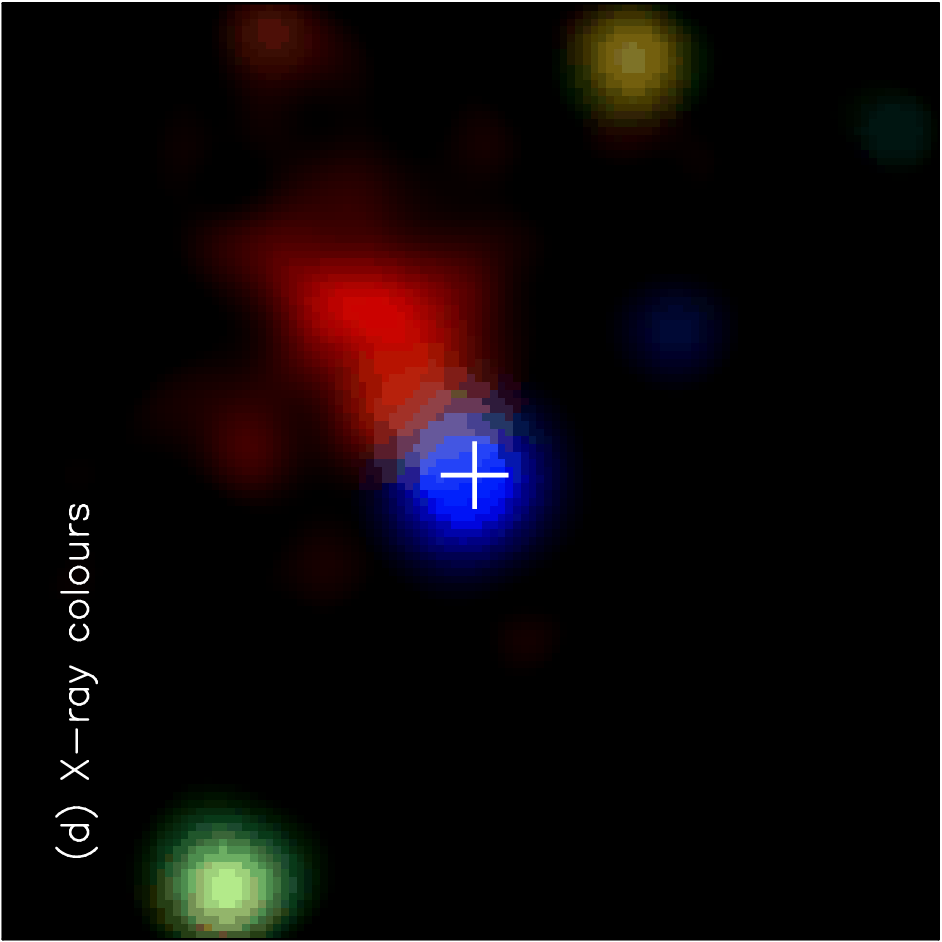}
}

\caption{ X-ray Images of the  nuclear region of NGC 4945. {\it Panels
a-c}:  \cha X-ray  images corresponding  to (a)  the soft  (0.3-1 keV)
band; (b) the medium (1-2 keV)  band and (c) the hard (2-10 keV) band.
{\it  Panel  d}:  \xmm  `X-ray  colour' image.  Red,  green  and  blue
correspond to the soft, medium and hard X-ray bands defined above. All
the images are  2\arcm~ on a side. The \cha  images have been smoothed
with an  adaptive `boxcar'  filter set to  encompass $\geq$10  cts per
box,  or to  have a  half-width of  5 pixels.   Non-adaptive, Gaussian
smoothing (2\arcs~ HWHM) was applied to the \xmm image.}  }
\label{4945CoreColour}
\end{minipage}
\end{figure}

\section{Joint spectroscopy with {\it CHANDRA} and {\it XMM-NEWTON}}

The  high  spatial resolution  afforded  by  \cha  has allowed  us  to
spatially resolve  several of the major components  that comprise the
total  X-ray  emission from  the  nucleus  of  NGC 4945.  However,  the
relatively  low  count-rate  means  the \cha  data  cannot be used to
place tight constraints   on    the   spectral    form   of   these
components. Conversely, despite the  greater count-rate, the lower
angular  resolution of \xmm  gives rise to some spatial
confusion, which complicates the identification and
spectral modelling of the components of the nuclear region. Fortunately, the  
{\it joint} analysis of the \cha and \xmm  spectra  removes  many  of 
these limitations.

We have adopted the following approach. We extract separate \cha ACIS spectra 
for the AGN and the extended emission components and model these 
in turn. The identified spectral components are then combined into a 
composite spectral model which is fitted to  the high  signal-to-noise 
\xmm EPIC  spectra of  the whole nuclear  region. Finally, we consider
combined fits to both the \cha and \xmm  spectra. As expected,
the latter provides  the tightest constraints on the individual model 
parameters.

\subsection{The \cha AGN spectrum}

The  spectrum  of  the point-like nuclear source (which we associate 
predominantly with the AGN in NGC 4945)  was  derived  from  the  
\cha GTI-corrected event list using a circular extraction region of
radius 6 pixels (3\arcs)  centred on the peak
of the X-ray  emission, at 13$^{h}$05$^{m}$27.5$^{s}$, -49\deg 28\arcm
05\arcs (J2000). The  corresponding background  spectrum was taken  from a
local source-free  region. The resulting spectrum
contains $\sim$750  counts (after  background subtraction) with no
significant signal below  1.5 keV. 

In modelling the spectral  form of the AGN 
we note that  in several well studied Seyfert  2 galaxies, with column
densities in excess  of a few $ \times  10^{23} \rm atoms~cm^{-2}$, it
appears  that  a  relatively unabsorbed  Compton-reflection  component
dominates the 2--10  keV spectrum (e.g. Mrk 3 -  Cappi \etal 1999, Sako
\etal  2000;  NGC   3281  -  Vignali  2002).  In   such  sources,  the
putative molecular torus may serve as a reprocessing site for the 
hard continuum generated by  the AGN. Furthermore, for a highly inclined 
configuration, it is perfectly plausible that the geometry is such as to 
give a relatively clear view of the far-side wall of the torus. It follows
that the observed Compton-reflection in Seyfert 2's may originated 
from this ``far-wall'' location. 

We have  tested this hypothesis
for  NGC  4945 by  fitting  the \cha  AGN  spectrum  with a  model
consisting of a very  heavily absorbed power-law continuum, unabsorbed
Compton  reflection of  the continuum  ({\it pexrav}  model  in XSPEC;
Magdziarz   \&  Zdziarski   1995)  and   an iron  K$\alpha$
fluorescence line  at 6.4 keV. It  turns out that  the underlying hard
continuum in  this model  has little impact  even at 10  keV, implying
neither the  \cha or \xmm  data can provide useful  constraints on
the parameters of the underlying continuum.  Thus, the continuum 
characteristics (photon
index,  $\Gamma$,   normalisation,  K$_{pl}$  and   absorbing  column,
$N_{H}$) were fixed at the values consistent with previous analyses of
the  high  energy  ($>$10  keV)  spectrum  from  both  \sax  and  \xte.
Specifically we take $\Gamma$=1.55,  K$_{pl}$=0.1 photon s$^{-1}$  
cm$^{-2}$ keV$^{-1}$ and $N_{H}$=4$\times$10$^{24}$~cm$^{-2}$ (Guainazzi  
\etal 2000; Madejski
\etal  2000). Also, the metal abundances were fixed at solar values, the
high-energy spectral cutoff set at  200 keV and the  relative reflection
parameter (R) was initially constrained to lie in the range 0--2 with 
cos({\it i})$=0.5$.

\begin{figure}
\centering
\begin{minipage}{85 mm} 
\centering
\vbox{ 
\includegraphics[width=5.5 cm, angle=270]{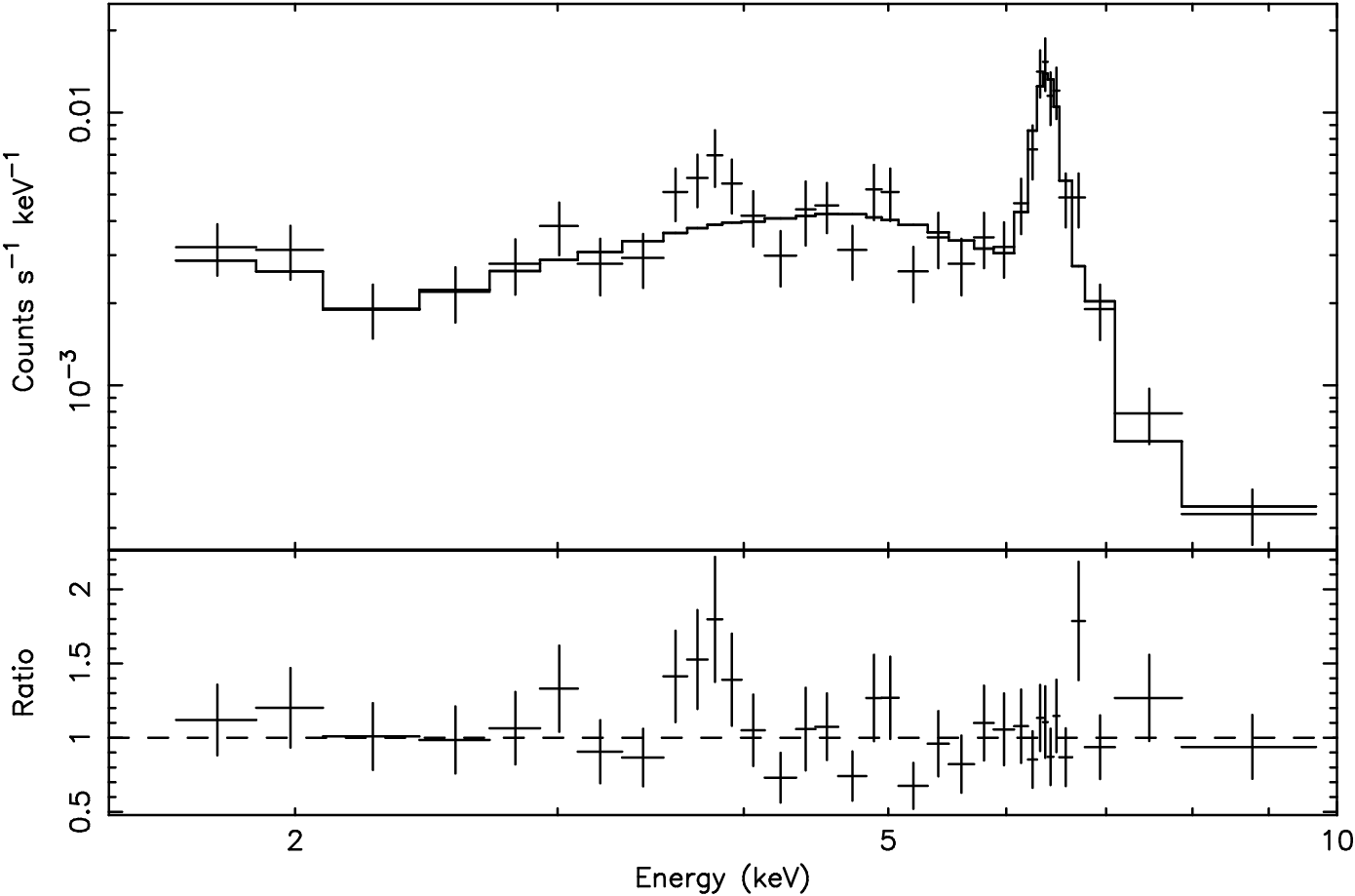}
\includegraphics[width=5.5 cm, angle=270]{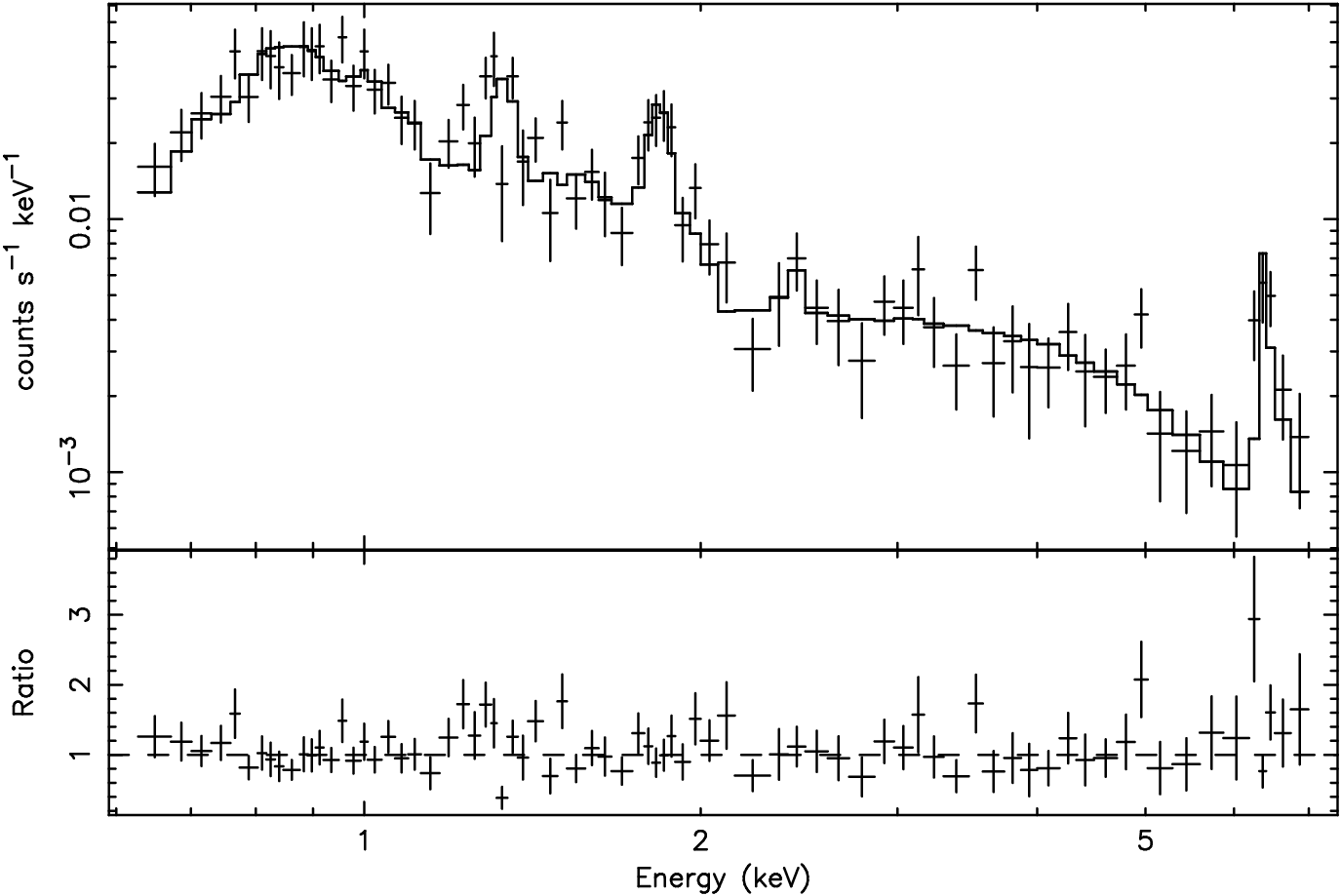}
}
\caption{{\it  Upper  Panel}:  \cha ACIS-S spectrum of the AGN in  NGC 4945.
The spectral  fit corresponds  to Model  A. The  data/model  residuals are
plotted below  the spectrum. {\it  Lower Panel}: \cha ACIS-S spectrum  of the
starburst/X-ray plume. The spectrum is  fitted with Model B and the
data/model residuals are again shown.}
\label{Chan_spect}
\end{minipage}
\end{figure}

This model,  with both the R parameter and the  energy, intrinsic width
and  normalisation  of  the  iron  K$\alpha$  line  as  free
parameters,  provides a  good fit  to  the \cha spectrum ($\chi^{2}$=32 for 29
degrees of freedom; hereafter d.o.f) as illustrated in  
Fig.  \ref{Chan_spect} (Upper Panel).  The best-fit
parameter values are given in Table \ref{models} (Model A), where the quoted
errors correspond  to 90\%  confidence levels  for one
interesting parameter ({\it i.e.}  $\Delta\chi^{2}$=2.71). Modelling the \cha
spectrum in  terms of a heavily absorbed direct continuum component
(but with $N_H$ an order of magnitude smaller than assumed above) 
plus a relatively unabsorbed
scattered power-law component ({\it cf.} Guainazzi   \etal   2000)  
yields a considerably worse fit ($\chi^{2}$=51.1 for 29 d.o.f). 
The fit residuals (Fig. \ref{Chan_spect}, Upper Panel) hint at 
some contamination of the \cha AGN spectrum by a nuclear starburst component;
our modelling of the composite spectrum takes account of this
in a straightforward fashion (see \S4.3).

\begin{table}
\scriptsize
\caption{Best-fit model parameters}
\begin{minipage}{65 mm}
\centering
\begin{tabular}{lcccl}
\hline
\multicolumn{5}{c} {} \\
Parameter & Model A & Model B & Model C & \\ 
\hline
R & 0.029$^{+0.003}_{-0.003}$ & - & 0.016$^{+0.002}_{-0.002}$ & \\ 
E$_{Fe k\alpha}$ & 6.42$^{+0.03}_{-0.03}$ & 6.4 $^{a}$ & 6.396 $^{+0.005}_{-0.007}$ & $^{b}$ \\ 
$\sigma_{Fe k\alpha}$ & 0.11$^{+0.04}_{-0.03}$ & 0 $^{a}$ & 0.048$^{+0.011}_{-0.013}$ & $^{b}$ \\ 
K$_{Fe k\alpha}$ & 2.2$^{+0.4}_{-0.4}\times10^{-5}$ & 0.7$^{+0.2}_{-0.3}\times10^{-5}$ & 2.4$^{+0.2}_{-0.2}\times10^{-5}$ & $^{c}$ \\ 
N$_{H, 1}$ & - & 0.16$^{a,d}$ & 0.16$^{a,d}$ & $^{e}$ \\
kT$_{1}$ & - & 0.61$^{+0.4}_{-0.9}$ & 0.60$^{+0.03}_{-0.03}$ & $^{b}$ \\
K$_{1}$ & - & 2.7$^{+0.4}_{-0.5}\times10^{-5}$ & 2.6$^{+0.3}_{-0.3}\times10^{-5}$ & $^{f}$ \\
N$_{H, 2}$ & - & 1.5$^{+0.1}_{-0.3}$ & 1.8$^{+0.1}_{-0.2}$ & $^{e}$ \\
kT$_{2}$ & - & 0.9$^{+0.1}_{-0.1}$ & 0.87$^{+0.08}_{-0.08}$ & $^{b}$ \\ 
K$_{2}$ & - & 5.1$^{+1.0}_{-1.2}\times10^{-4}$ & 6.5$^{+0.8}_{-1.1}\times10^{-4}$ & $^{f}$ \\ 
N$_{H, 3}$ & - & 8.8$^{+4.4}_{-3.5}$ & 12.1$^{+1.1}_{-1.6}$ & $^{e}$ \\ 
kT$_{3}$& - & 8.7$^{g}$ & 6.0$^{+1.1}_{-0.8}$ & $^{b}$ \\ 
K$_{3}$ & - & 4.5$^{+1.6}_{-1.2}\times10^{-4}$ & 6.4$^{+1.1}_{-0.9}\times10^{-4}$ & $^{f}$ \\ \hline
$\chi^{2}$/d.o.f & 32.3/29 & 81/67 & 290/269 & \\ \hline
\multicolumn{2}{l}{$^{a}$ Fixed parameter} & \multicolumn{2}{l}{$^{b}$ keV} \\
\multicolumn{2}{l}{$^{c}$ photons s$^{-1}$ cm$^{-2}$} & \multicolumn{2}{l}{$^{d}$ Galactic column density} \\
\multicolumn{2}{l}{$^{e}$ $\times10^{22}$ cm$^{-2}$} & \multicolumn{2}{l}{$^{f}$ photons s$^{-1}$ cm$^{-2}$ keV$^{-1}$} \\
\multicolumn{2}{l}{$^{g}$ Unconstrained parameter} & \\
\end{tabular}
\end{minipage}
\label{models}
\end{table}

\subsection{The \cha starburst/X-ray plume spectrum}

The  \cha spectrum of the extended nuclear X-ray emission
encompassing the resolved nuclear starburst component and the X-ray plume 
structure was constructed using an elliptical extraction region
centred  on  13$^{h}$05$^{m}$25.5$^{s}$,
-49\deg 27\arcm 54\arcs (J2000), with the major axis aligned along 
a position  angle of $120^{\circ}$, with a semi-major  axis of 30.5\arcs
and a semi-minor axis  of 13.4\arcs.  In this case, the AGN component 
was suppressed by excluding the region within 3\arcs of the nucleus. 
The  resulting spectrum (after background subtraction
using a local source-free region) contains  $\sim$1600  counts.  \cha is  
capable of resolving  the starburst region  from  the X-ray  plume,  
but  here we have chosen to utilize  the  better signal-to-noise  of the 
combined spectrum. Taking as  a lead the published results for the starburst 
and  X-ray plume in NGC 253 (Pietsch \etal 2001), we have modelled the 
\cha  spectrum  with three solar abundance thermal ({\it mekal})
components, each
exhibiting  different  levels  of  absorption  as well as different
temperatures and  normalisations. 

This model yields a  good fit to
most of the observed spectral features, including the  significant
K$\alpha$ emission from helium-like iron at 6.7 keV. However it does not 
account for the K$\alpha$ emission from neutral iron observed at 6.4 keV.
Analysis of the
individual  starburst  and  X-ray  plume  spectra  confirm  that  both
K$\alpha$ emission  lines originate
entirely  from  the  starburst  region.  We  model  the  neutral  iron
K$\alpha$  emission  with  the  addition of  an  intrinsically  narrow
Gaussian line.  A  model comprising three thermal components 
plus an iron K$\alpha$ line provides a reasonable fit to  the data ($\chi^{2}$=89
for 67 d.o.f, see Fig. \ref{Chan_spect}; {\it Lower Panel}). The corresponding
best-fit model parameters are listed in Table \ref{models} (Model B).

\subsection{Joint fitting of the composite \xmm and \cha spectra}

The \xmm spectra for the combined AGN, starburst and X-ray 
plume components were extracted using a 30\arcs~radius circle centered
on  13$^{h}$05$^{m}$28$^{s}$, -49\deg  28\arcm 01\arcs (J2000).  For  both the
MOS and  PN cameras, a corresponding
background spectrum  was extracted from  a 90\arcs~radius circle  in a
source  free  region of  the  same  CCD.  After background subtraction,
each MOS spectrum  contained $\sim$1350 counts compared to 
$\sim$3500 counts in the PN spectrum.  


The \xmm spectrum was analysed by combining the two models 
used in the description of the \cha  observations (Models A \& B in
\S4.1 \& \S4.2). This composite model yields an  excellent fit to the
broadband \xmm spectra ($\chi^{2}$=273.5  for 268 d.o.f). However,
an  unphysical trade-off between  the  hottest thermal  component  
and  the Compton-reflection parameter result in a value of R  
which is inconsistent with the value found from the \cha AGN spectrum 
alone (assuming that  this component has not  varied significantly 
over  the one-year  interval  between the  two  observations, see  \S 5.2).  
We have  addressed this  problem by  performing  {\it joint spectral fitting} 
of three spectral datasets, namely the \xmm composite  spectrum and the 
\cha AGN and starburst/X-ray plume spectra.
Only the Compton-reflection 
component, the iron line and  the  hottest  thermal  
component were utilized in fitting the \cha AGN spectrum, whilst
only the three thermal components and the iron line were employed
in relation to  the \cha  starburst/X-ray  plume  spectrum. We also
adapted the spectral model somewhat by fixing the absorption applied to the 
softest thermal component at the Galactic value and setting the  absorption on
the Compton-reflection component to be the same as that
applied to the intermediate-temperature thermal component  
({\it i.e.} $N_{H,2}$). 

\begin{figure}
\centering
\begin{minipage}{85 mm} 
\centering
\vbox{ 
\includegraphics[width=5.5 cm, angle=270]{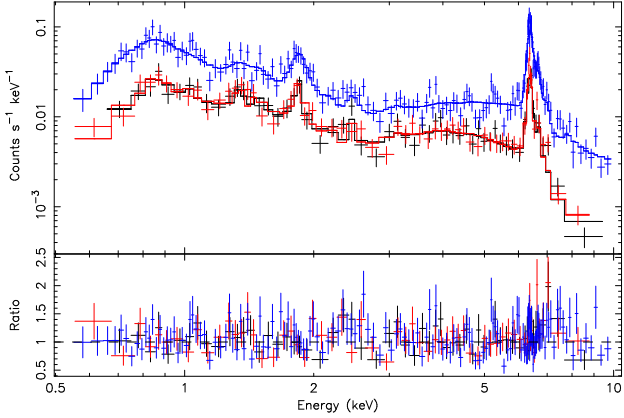}
}
\caption{\xmm EPIC  MOS (black and red)  and PN (blue)  spectra of the
extended nuclear region  of NGC  4945. The spectra  are fit with  the composite
model (Model  C) and  the data/model residuals  are also shown.}
\label{xmm_spect}
\end{minipage}
\end{figure}

Simultaneous fitting of the composite spectral model described above
gives a good match to all three spectral datasets. The resulting 
best-fit  to the \xmm data is shown in  Fig. \ref{xmm_spect}; 
the corresponding parameters are given in Table \ref{models} (Model C). 
The high signal-to-noise of
the  \xmm spectra,  relative to  the  \cha spectra, leads to
tighter constraints on  all the  parameters. The  
simultaneous fits highlight  the  important  contribution  of the  
highest  temperature thermal  (starburst)  component to  the 0.5-10  keV  
spectrum of  NGC 4945. In this respect the  joint spectral  fitting  is 
particularly valuable since the \cha starburst/outflow spectrum has 
insufficient counts to put useful limits on the temperature  of 
this component.

When data relating solely to the  spatially-resolved X-ray  plume  are 
extracted (from both the \cha and \xmm datasets), the spectrum is found  
to be extremely soft with few counts above 2 keV. Spectral fitting
using a single temperature {\it mekal} model, modified by Galactic absorption,
gives  a  temperature  and  normalisation  completely consistent  
with those of the softest  thermal  component in the composite  
spectral model (Model  C), identifying  this component  with the X-ray plume emission.  
The two remaining thermal  components  are attributable   
to  the  nuclear starburst.

\begin{figure}
\centering
\begin{minipage}{85 mm} 
\centering
\vbox{ 
\includegraphics[width=6.1 cm, angle=270]{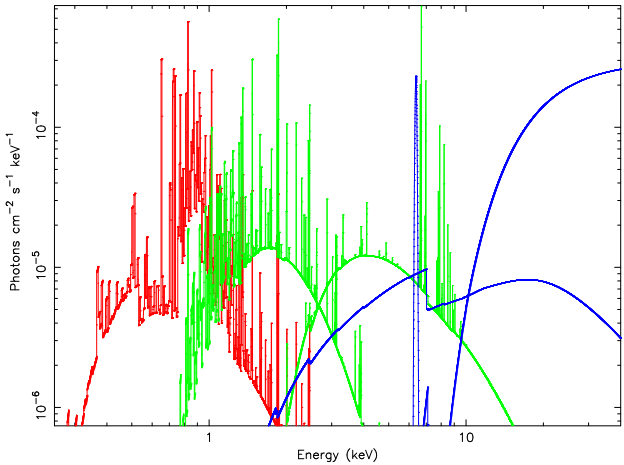}
}
\caption{The broadband spectral model for the nucleus of NGC 4945. The
component  associated  with the  X-ray  plume  is  shown in  red,  the
starburst components in green and the AGN components in blue.}
\label{model}
\end{minipage}
\end{figure}

\section{Discussion}

The components of the broadband  spectrum  of the extended nucleus of NGC
4945 are shown in Fig. \ref{model}. The Compton-reflection component
makes  a significant  contribution between  5 and 10 keV,
whereas at  lower energies  the spectrum is  dominated by  the thermal
emission from the nuclear starburst and X-ray plume. The  heavily absorbed 
AGN continuum  only emerges above 10 keV. Strong  
K$\alpha$ emission lines  from neutral and helium-like iron are also 
present in the spectrum at 6.4 and 6.7 keV respectively. 

\subsection{The reprocessing of the AGN continuum}

In the  unified model proposed  by Antonucci (1993), when an active nucleus 
is viewed at relative high inclination, the line of sight may intercept
the surrounding molecular torus and hence be subject to significant
obscuration. Previous observations testify to a near edge-on configuration 
of the inner disk in NGC 4945 (Greenhill, Moran \&  Hernstein 1997) 
and a very high line-of-sight column density to its active nucleus
(Guainazzi \etal 2000; Madejski \etal 2000).
Consistent with this picture, the analysis of the \xmm  and \cha spectra 
of the nuclear region of  NGC 4945  demonstrates  that in  the  
0.5-10 keV  range, there is only indirect evidence for the presence of the 
AGN (since the direct continuum is strongly cut-off below $\sim 10$ keV).

One signature of the underlying AGN in NGC 4945 is the Compton-reflection 
continuum which we attribute to reprocessing off the far, inside wall of 
the molecular torus. In this scenario the ratio of the reflected
to the direct continuum as embodied in Compton-reflection parameter, R, 
can be used to place a lower limit on the inclination angle of the torus, 
provided the geometrical configuration of the absorbing torus is known. In 
reality this is very uncertain,  although the half-opening
angle of tori are typically in the range $20^{\circ}-50^{\circ}$
({\it e.g.} Wilson \& Tsvetanov 1994). If we assume the geometry 
envisaged by Ghisellini, Haardt \& Matt (1994), then our estimate of
$R=0.016$ (based on the `average' continuum given in Guainazzi \etal 2000)
implies a near edge-on view of the torus. Alternatively a near edge-on but 
geometrically thin (half-angle <10$^{o}$) distribution of absorbing material 
located on parsec scales, as inferred by Madejski \etal (2000) on the basis 
of the hard X-ray variability seen in NGC 4945, is also consistent with a 
very low value for R.

The neutral iron K$\alpha$ line  is narrow and the nuclear component
has an equivalent width of 1.6$\pm$0.1 keV with  respect to the observed 
Compton-reflection continuum. This is fully consistent with the line and the 
reflected continuum originating at the same location, namely the visible 
part of the torus wall ({\it e.g.} Matt, Brandt \& Fabian 1996).
There is also a $30\%$ contribution to the neutral iron 
K$\alpha$ emission from  the extended nuclear  starburst region, implying  
that  cool  material in  the environment of the starburst is  also exposed 
to the AGN continuum  flux. 

Although the hard  continuum emanating from the Seyfert  2 nucleus of
NGC  4945  is  known  to  be  variable  on  timescales  as  short  as
$\sim10^{4}$s (Guainazzi  \etal 2000), no  significant variability is
detected in either the \cha or \xmm observations, nor is 
it evident from the joint spectral fits (the time interval between the 
two sets of observations being 12 months). This tallies with the location
of the continuum reprocessing, namely the torus wall, lying at a distance
in excess of a light year from the nuclear continuum source. 
(The apparent discrepancy between the \cha-only and the joint-fit estimates
of R is due to the fact that the initial  modelling of the \cha AGN spectrum 
did not account for the starburst contamination of the AGN spectrum).

\subsection{The nuclear starburst}

The spectral modelling discussed above characterises the nuclear starburst 
in terms of two thermal components with temperatures of  $\sim 0.9$
and $\sim 6.0$ keV. Very similar temperatures were observed (at least 
in terms of the hottest thermal components) in the 
nearby starburst galaxy NGC 253 by Pietsch \etal (2001),  
suggesting that  the presence  of  the AGN in NGC 4945
has  little impact  on  the  thermal  properties  of its nuclear
starburst (there is no evidence for  an AGN in NGC 253).  Gross
features, associated with the cooler of the two starburst components,
include the emission lines of Mg X/XI (1.34 keV), Si XII/XIII  (1.83 keV) 
and S XIV/XV (2.45  keV). At higher energy, strong K$\alpha$  line emission 
from helium-like iron is attributable to the hotter starburst component.
High temperature plasma  and associated line emission  is  
characteristic of  X-ray emission from  a population of type  Ib and 
type IIa  SNR's (e.g Behar \etal 2001). No  corresponding iron edge  
is detected in either  of the \cha or \xmm observations.  The unabsorbed 
luminosity of the starburst region measured  from \xmm (0.5-10 keV)  
is $\sim 8 \times10^{39}$ erg s$^{-1}$. Analysis  of the separate starburst 
and  X-ray plume spectra shows  that the  nuclear  starburst emission  
declines rapidly below $\sim 1.5$ keV, probably as a result of absorbing 
material in the plane of the galaxy. This absorption may originate in 
material forming a low scale height disk and/or in material associated 
with the optical dust lanes of the galaxy.

\subsection{The X-ray plume}

Recent \cha (Strickland et al.  2000) and \xmm (Pietsch et  al. 2001) 
observations have established that the outflow driven by the nuclear
starburst in NGC 253 has a limb-brightened conical X-ray morphology. 
NGC 4945 provides a second  example of such a structure. In both 
galaxies the limb-brightened X-ray structure correlates well with
the observed the H$\alpha$ emission.  A similar X-ray temperature
is derived in each case. (In NGC 253 a single temperature thermal 
plasma with $kT \sim 0.6$ keV  fits  the  \cha  spectrum reasonably
well,  although  the  situation  is  more complicated in  the better 
quality  \xmm data.)  However, there  are also significant  differences.  
For  example, the  NGC 4945 X-ray plume displays  a wider opening angle 
($\theta  \sim 40^{\circ}$), and the limb  brightening is most evident  
in the 1--2 keV, band whereas in NGC 253  both  the  \cha and  \xmm  
data show  the limb  brightened structure is apparent down to 0.5  keV.  
A similar  physical model to that postulated by Pietsch et al.  (2001) 
can be inferred for NGC  4945.  The limb-brightened  structure can 
be  attributed  to highly-excited  gas with  a low volume-filling  factor,
which is produced by an   interaction  between  the
starburst-driven wind  and the dense ISM surrounding  the outflow. In
the case of NGC 4945, the comparative uniformity of the emission  
below  1  keV may  be a  direct
observation  of a  mass-loaded  superwind emanating  from the  nuclear
starburst (as described  by Suchkov \etal 1996).  The  question of why
the limb-brightened component  is hotter in NGC 4945 than in  NGC 253 
is  an interesting point  for future,  detailed studies  of  this system.   
We derive  an
intrinsic luminosity of $\sim1.1\times 10^{38} \rm ~erg~s^{-1}$ (0.5 -
2 keV) for the superwind component from the \xmm data.

\section{Conclusions}

We present a detailed imaging and spectroscopic analysis of the
nuclear region of NGC 4945 using data from both \cha and \xmm.  The
main results of our study are:

\begin{itemize}
\item
The nuclear regions display a complex X-ray morphology, including a
predominantly hard source at the nuclear position (which is partially
resolved in the \cha data) and a soft X-ray ``plume'' extending to the
northwest of the nucleus.
\item
The 0.5-10 keV emission from the nuclear source is dominated by a
combination of reprocessed direct AGN continuum (in the form of
neutral Compton reflection and iron K$\alpha$ line emission) and
thermal emission from a nuclear starburst.  The underlying AGN continuum
is not observed directly in either the \cha or \xmm data, due to heavy
($\sim4\times10^{24}$ cm$^{-2}$) absorption in the line of sight to
the continuum source, that cuts off the continuum below $\sim$10 keV.
\item
The neutral iron K$\alpha$ line is well fit with a narrow Gaussian
profile (with an equivalent width of 1.6$\pm$0.1 keV), consistent with
the line and the reflected continuum originating from the visible part
of the torus wall.  Furthermore, the best-fit value of the relative
reflection parameter, $R=0.016$, implies a near edge-on geometry for
the torus.
\item
Neutral iron K$\alpha$ emission from the extended nuclear starburst
region, identified with \cha, implies that cool material in the
environment of the starburst is also exposed to the AGN continuum
flux. The thermal emission from the nuclear starburst is fit with a
two temperature model, with $kT \sim$0.9 and $\sim$6.9 keV. The
spectrum shows clear features associated with the two starburst
components, including emission lines of Mg X/XI, Si XII/XIII and S
XIV/XV from the cooler component and strong K$\alpha$ line emission
from helium-like iron originating in the hotter component.
\item
A \cha image reveals the X-ray plume to have a limb-brightened
morphology in the 1--2 keV band. The plume is associated with the
softest emission observed by \xmm and is well modelled by a single
temperature thermal model, with $kT \sim$0.6 keV. When combined, these
properties suggest a physical interpretation of the plume as a
mass-loaded superwind emanating from the nuclear starburst.
\end{itemize}

Finally, we note that combining the  \cha and \xmm datasets 
has helped to mitigate the limitations of the individual observations
and allowed us to carry out a much more detailed imaging and spectral 
study than would otherwise have been possible.  This work clearly 
demonstrates the potential of joint \cha and \xmm analysis in characterising 
spatially and spectrally complex X-ray sources, such as those associated 
with the extended nuclear regions of nearby galaxies.

\vspace{1cm}

{\noindent \bf ACKNOWLEDGMENTS}

\vspace{2mm}

NJS gratefully acknowledges the financial support from PPARC. It is a 
pleasure to thank Kevin Briggs and Dick Willingale for their valuable 
aid with the Q software system. We also wish to thank the \xmm and \cha 
instrument and calibration teams for their on-going efforts to fully 
characterise and calibrate their respective instruments. The archival \cha 
data were obtained from the UK mirror of the \cha X-ray Center Data Archive 
operated by the Leicester Data Archive Service (LEDAS). This research has 
made extensive use of NASA's Astrophysics Data System Abstract Service.


\end{document}